\documentstyle[12pt]{article} 
\newcommand{\ii}{\'{\i}}

\newcommand{\zero}{\setcounter{equation}{0}}

\begin{document}

\title{Superconducting cosmic string with propagating torsion }

\author{C.N. Ferreira$^1$ \thanks{crisnfer@cbpf.br}, H.J. Mosquera
Cuesta$^{1,2}$ \thanks{herman@ictp.trieste.it} \\
and \\  L.C. Garcia de
Andrade$^{3,4}$ \thanks{garcia@dft.if.uerj.br}}

\date{}

\maketitle

\begin{center}

$^1$Centro Brasileiro de Pesquisas F\'{\i}sicas, Rua Dr.
Xavier Sigaud 150, Urca 22290-180, Rio de Janeiro,RJ, Brazil

$^2$ Abdus Salam International Centre for Theoretical Physics, High Energy, 
 Strada Costiera 11, Miramare 34014, Trieste, Italy
  
$^3$ Departamento de F\'{\i}sica Te\'orica-Instituto de F\'{\i}sica-UERJ, 
Rua S\~ao Francisco Xavier 524, 20550-003, Rio de Janeiro Maracan\~a Brazil,

$^4$ Departamento de Matem\'atica Aplicada-IMEECC, Universidade Estadual de Campinas, 13081-970 
Campinas/SP, Brazil

\end{center}

\begin{abstract} We show that it is possible to construct a consistent
model describing a current-carrying cosmic string endowed with torsion.  
The torsion contribution to
the gravitational force and geodesics of a test-particle moving around
the SCCS are analyzed. In particular, we point out two interesting
astrophysical phenomena in which the higher magnitude force we derived may
 play a critical role:
the dynamics of compact objects orbiting the torsioned SCCS and accretion of
matter onto it. The deficit angle associated to the SCCS
can be obtained and compared with data from the Cosmic Background
Explorer (COBE) satellite. We also derived a value for the torsion
contribution to matter density fluctuations in the early Universe.
\end{abstract}

\newpage

\vspace{.5 true cm}

\section{Introduction}

Cosmic strings have exact solutions \cite{Vilenkin} which
represent topological defects that may have been formed during
phase transitions in the realm of the Early Universe
\cite{Kibble}. The GUT defects carry a large energy density and
hence are of interest in Cosmology as potential sources for
primordial density perturbations. These fluctuations would leave
their imprint in the cosmic microwave background radiation (CMBR);
a prediction not ruled out by COBE satellite
observations yet\cite{smoot99}, and hence would act as seeds for
structure formation and thus as builders of the largest-scale
structures in the Universe\cite{Brandenberger}, such as
the very high redshift superclusters of galaxies as for instance
the {\it great wall}. They may also help to explain the most
energetic events in the Universe such as the cosmological gamma-ray bursts
(GRBs)\cite{brandeb93}, ultra high energy
cosmic rays (UHECRs) and very high energy
neutrinos\cite{Brandenberger} and gravitational-wave
bursts and backgrounds\cite{Allen94}. All these are
issues deserving continuous investigation by many physicists
nowadays \cite{Hawking}.

 Witten \cite{Witten1} has shown that the cosmic strings may possess
superconducting properties and may behave like
  bosonic (see Ref.[\cite{Patrick1}] and references therein) or
fermionic strings\cite{Davis99}. In other communications it was supposed
  that the relevant superconductivity is generated during or very
  soon after the primary phase transition in which the string is
formed.  Cartan
torsion has been connected previously with ordinary cosmic strings
\cite{Soleng,Garcia1} and also spinning cosmic strings
\cite{Letelier1} from quite distinct point of views.

In this work we consider the study case of bosonic SCCSs in  
Riemann-Cartan space-time with coupling terms in the potential. One should regard such
an extension as a first step of a comprehensive study of
 cosmic string models in the context of theories including torsion \cite{Herman1}. We
aim at dealing with most realistic models which demand
 supersymmetry, an essential ingredient of grand-unification theories,
string theory, etc. Thus, we ought to combine both gravitational and
spin degrees of freedom in the same formalism, and thus torsion is required.

The main-stream of this paper is as follows: we explored the physics of
torsion coupling to cosmic strings in section II. An external solution
 for the SCCS metric in this scenario in presented in section III,
while in section IV we derive the corresponding one for the internal
structure of the SCCS by using the weak-field approximation. Two applications
are provided. One focusing on the deviation of a particle moving near
a torsion string. It is shown that such a high intensity of the gravitational
force from the screwed SCCS (when compared with the one generated by a current-carrying
 string) may have important effects on the dynamics of
compact objects orbiting around it, and also on matter being
accreted by the string itself. The second one exploits the
possibility that the temperature fluctuations in the cosmic
 microwave background radiation could have been, at least
 partially, generated after the SCCSs having interacted with it.
 We obtain a neat expression for the deficit angle in this context
 and a comparison is done with data from COBE satellite.
 We end this paper with a short summary of the picture here suggested.

\section{Torsion coupling to cosmic strings}

Here we construct a consistent framework for the torsion field 
pervading a cosmic string
and define the vortex configuration for this problem. We choose here
to analyse the simplest case where the torsion appears. In this line
of reasoning, it is possible to describe torsion as a gradient-like
field \cite{Kleinert}

\begin{equation} S_{\mu \nu}^{\hspace{.3 true cm} \lambda}
=\frac{1}{2} [\delta_{\mu}^{\lambda} \partial_{\nu} \Lambda -
\delta_{\nu}^{\lambda}\partial_{\mu} \Lambda ], \label{torc1}
\end{equation}

\noindent
being the $\Lambda$ field the source of torsion in the string that have only 
r-dependence. 

The action representing
 the SCCS in a space-time with torsion can be
written as:

\begin{equation} S = \int d^4x \sqrt{g} \left[\frac{1}{16 \pi
G}R(\{\}) + \alpha_1 \nabla_{\mu}S^{\mu} + \alpha_2 S_{\mu \nu
k}S^{\mu \nu k} + \alpha_3 S_{\mu \nu k}S^{\mu k \nu} + \alpha_4
S_{\mu}S^{\mu}\right] + S_m, \label{acao1} \end{equation}

\noindent
where $R(\{\})$ is the curvature scalar of the Riemannian theory and
$S_m$ is the matter action that describes the superconducting cosmic
string (to be specified below). Here $\alpha_1$, $\alpha_2$, $\alpha_3$ and 
$\alpha_4$ are coupling 
arbitrary constants with $\alpha_1$ connected
with the torsion gradient term $\nabla_{\mu} S^{\mu}$. Where
$\nabla_{\mu}$ is a Riemannian covariant derivative which drops out
from the action because it is a term involving a total derivative.
$S_{\mu \nu k}$ and $S_{\mu}$ are $SO(1,3)$
irreducible components of the torsion. For this extended Riemann-Cartan  
space (see Ref.[\cite{Shapiro1}] for a review), the affine connection can be written in terms of $g_{\mu \nu}$ and
$S_{\alpha} = \partial_{\alpha} \Lambda $ as

\begin{equation}
 \Gamma_{\lambda \nu}^{\hspace{.3 true cm} \alpha} =
\{^{\alpha}_{\lambda \nu}\} +
 S^{\alpha}g_{\lambda \nu} - S_{\lambda} \delta^{\alpha}_{\nu} \label{kont1}
\end{equation}

\noindent
so that $S_{\mu}$ is the only piece that contributes to torsion,
which here is the escalar derivative defined by Eq.(\ref{torc1}).

Then we may consider a theory of {\it gravitation} possessing
torsion by writing that part of the action $S_G$ stemming from the
curvature scalar $R$ as:

\begin{equation} S_G = \int d^4x \sqrt{g} \left[\frac{1}{16 \pi
G}R(\{\}) - \frac{\alpha}{2} \partial_{\mu} \Lambda
\partial^{\mu} \Lambda \right],
\end{equation}

\noindent
where the coupling constant $\alpha $ is 
related with $\alpha_2$, $\alpha_3$ and $\alpha_4$ and will be specified
with the help of COBE data.

 We can study the SCCS considering the Abelian Higgs
 model with two scalar fields, $\phi$ and $\tilde \Sigma $. In this case,
the action for all matter fields turns out to be:

\begin{equation} S_m = \int d^4x \sqrt{{g}}[-\frac{1}{2}D_{\mu}\phi
(D^{\mu}\phi)^* - \frac{1}{2}D_{\mu} \Sigma (D^{\mu}
\Sigma)^* - \frac{1}{4}F_{\mu \nu}F^{\mu \nu} - \frac{1}{4}H_{\mu
\nu}H^{\mu \nu} - V(|\phi|, |\Sigma |, \Lambda)],
\end{equation}

\noindent
where $ D_{\mu} \Sigma = (\partial_{\mu} + ie A_{\mu})
\Sigma $ and $D_{\mu} \phi = ( \partial_{\mu} + i
qC_{\mu})\phi $ are the covariant derivatives. The reason why the gauge fields do not minimally 
couple to torsion is well discussed in the works of references \cite{Gaspperini,Hehl}.  
The field strenghts are defined as usually as 
$ F_{\mu \nu} = \partial_{\mu}A_{\nu} - \partial_{\nu} A_{\mu}$ and $ H_{\mu \nu} = \partial_{\mu}C_{\nu} -
\partial_{\nu} C_{\mu} $, with $A_\mu$ and $C_\nu$ being the gauge fields. 

The potential $V(\varphi, \sigma, \Lambda)$ triggering the
symmetry breaking can be fixed by:

\begin{equation} V(\varphi, \sigma, \Lambda) = \frac{\lambda_{\varphi}}{4} (
\varphi ^2 - \eta^2)^2 + f_{\varphi \sigma}\varphi ^2\sigma ^2 +
\frac{\lambda_{\sigma}}{4}\sigma ^4 -
\frac{m_{\sigma}^2}{2}\sigma^2 + l^2\sigma^2 \Lambda^2, \end{equation}

\noindent
where $\lambda_{\varphi}$, $\lambda_{\sigma}$, $f_{\varphi \sigma}$  and $l^2$  
are coupling constants, and the boson mass being defined by $m_{\sigma}$.

This action (Eq.\ref{acao1}) has a $U(1)' \times U(1)$ symmetry, where the
$U(1)' $ group, associated with the $\phi$-field, is broken by the vacuum and
gives rise to vortices of the Nielsen-Olesen\cite{Nielsen}

\begin{equation} \begin{array}{ll} \phi = \varphi(r )e^{i\theta},& \hspace{.5 true cm}
C_{\mu} = \frac{1}{q}[P(r) - 1]\delta^{\theta}_{\mu},
\end{array}\label{vortex1} \end{equation}

\noindent
parametrized in cylindrical coordinates $(t,r,\theta,z)$, where
$r\geq 0$ and $0 \leq \theta < 2 \pi $. 
The boundary conditions for the fields $\varphi(r) $ and
$P(r)$ are the same as those of ordinary cosmic
strings\cite{Nielsen}:

\begin{equation} \begin{array}{llll} \varphi(r) = \eta & r \rightarrow
\infty \hspace{.3 true cm} \varphi(r) =0 & r = 0 \end{array} \hspace{1
true cm}
\begin{array}{llll} P(r) =0 & r \rightarrow \infty &\hspace{.3 true cm} P(r) =1 & r= 0.  \end{array} \label{config1} \end{equation}

The other $U(1)$- symmetry, that we associate to electromagnetism, acts on the  $\Sigma $-field. This symmetry is not broken by the vacuum; however, it is broken in the interior of the deffect. The $\Sigma $-field in
the string core, where it
 acquires an expectation value, is responsible for a bosonic current
being carried by the gauge field $A_{\mu}$. The only non-vanishing
components of the gauge fields are $A_z(r)$ and $A_t(r)$ and the
current-carrier phase may be expressed as $\zeta(z,t) = \omega_1 t -
\omega_2z$. Notwithstanding, we focus only on the magnetic case
\cite{Patrick1}. Their configurations are defined as:

\begin{equation} \begin{array}{ll} 
\Sigma = \sigma(r)e^{i\zeta(z,t)},& \hspace{.3 true cm}
A_{\mu} = \frac{1}{e}[A(r) - \frac{\partial \zeta(z,t)}{\partial
z}]\delta_{\mu}^{z},  
\end{array} \label{vortex2} \end{equation}

\noindent
because of the rotational symmetry of the string itself. The fields
responsible for the cosmic string superconductivity have
the following boundary conditions:

\begin{equation} \begin{array}{llll} \frac{d}{d r}\sigma(r) = 0 & 
r=0 & \hspace{.3 true cm}\sigma(r) = 0 & r \rightarrow \infty \end{array}
\hspace{1 true cm} \begin{array}{llll} A(r) \neq 0 & r \rightarrow
\infty &\hspace{.3 true cm} A(r) = 1 & r = 0.
\end{array}\label{config2}
\end{equation}

 Let us consider a SCCS in a cylindrical coordinate system $(t,r,\theta,z)$, 
 so that $r \geq 0$ and $0 \leq \theta < 2 \pi $ with the
metric defined in these coordinates as:

\begin{equation} ds^2 = e^{2(\gamma - \psi)}(-dt^2 + dr^2 ) +
\beta^2e^{-2\psi}d\theta^2 + e^{2\psi}dz^2, \label{metric1}
\end{equation}

\noindent
where $\gamma, \psi$ and $\beta$ depend only on $r$. We can write Einstein-Cartan equations in
the quasi-Einsteinian form:

\begin{equation} G^{\mu}_{\nu}(\{\}) =8 \pi G (2 \alpha g^{\mu
\alpha}\partial_{\alpha} \Lambda \partial_{\nu} \Lambda  - \alpha
\delta^{\mu}_{\nu} g^{\alpha \beta}
\partial_{\alpha } \Lambda
\partial_{\beta } \Lambda  +
T^{\mu}_{\nu})  = 8\pi G \tilde T^{\mu}_{\nu} \label{gmg}
\end{equation}

\noindent
where $(\{\})$ stands for Riemannian geometric objects, 
$\delta^{\mu}_{\nu} $ and $T^{\mu}_{\nu}$ correspond to the
identity and energy-momentum tensors, respectively. 
$\tilde T^{\mu}_{\nu}$ tensor corresponds to an eneregy-momentum 
tensor containing the torsion field.

We have seen that the dependence upon torsion is represented, in the
quasi-Einstenian form, by the $\Lambda$-field that has an equation of
motion given by Eq.(\ref{phi}) below, whose solution shall be addressed
subsequently.

The SCCS energy-momentum tensor is defined by $T^{\mu}_{(scs)\nu} = \frac{2}{\sqrt{g}} \frac{\delta S_m}{\delta g_{\mu \nu}}$, which yields:

\begin{equation}
\begin{array}{ll}
T_{scs \hspace{.1 true cm}t}^t=& -\frac{1}{2} \left\{ e^{2(\psi -
\gamma)}[\varphi'^2 + \sigma'^2] +
\frac{e^{2\psi}}{\beta^2}\varphi^2P^2 + e^{-2\psi} \sigma^2 A^2 +
\right.\\
&\left.+
\frac{e^{2(2\psi-\gamma)}}{\beta^2}(\frac{P'}{q})^2 +
e^{-2\gamma}(\frac{A'}{e})^2 + 2V(\varphi,\sigma, \Lambda)\right\}
\end{array}
\label{tens1}
\end{equation}

\begin{equation}
\begin{array}{ll}
T_{scs \hspace{.1 true cm}r}^r =& \frac{1}{2} \left\{e^{2(\psi -
\gamma)} [\varphi'^2 + \sigma'^2] - \frac{e^{2\psi}}{\beta^2}
\varphi^2P^2 - e^{-2\psi}\sigma^2 A^2 +\right.\\
&\left.+
\frac{e^{2(\psi-\gamma)}}{\beta^2}(\frac{P'}{q})^2 +
e^{-2\gamma}(\frac{A'}{e})^2 - 2V(\varphi,\sigma, \Lambda)\}\right\}
\end{array}
\label{tens2}
\end{equation}

\begin{equation}
\begin{array}{ll}
T_{scs \hspace{.1 true cm}\theta}^{\theta}=& -\frac{1}{2} \left\{
e^{2(\psi - \gamma)} [\varphi'^2 + \sigma'^2] -
\frac{e^{2\psi}}{\beta^2}\varphi^2P^2 + e^{-2\psi }\sigma^2 A^2 +
\right.\\
& -
\left.\frac{e^{2(\psi-\gamma)}}{\beta^2}(\frac{P'}{q})^2 +
e^{-2\gamma}(\frac{A'}{e})^2 + 2V(\varphi,\sigma, \Lambda) \right\}
\end{array}
\label{tens3}
\end{equation}

\begin{equation}
\begin{array}{ll}
T_{scs \hspace{.1 true cm}z}^z =& -\frac{1}{2} \left\{ e^{2(\psi
- \gamma)}[\varphi'^2 + \sigma'^2] +
 \frac{e^{2\psi}}{\beta^2}\varphi^2P^2 - e^{-2\psi }\sigma^2 A^2 +\right.\\
&\left. +
\frac{e^{2(\psi-\gamma)}}{\beta^2}(\frac{P'}{q})^2 -
e^{-2\gamma}(\frac{A'}{e})^2 + 2V(\varphi,\sigma, \Lambda)\right\}.
\end{array}
\label{tens4}
\end{equation}

In these expressions Eqs.(\ref{tens1}-\ref{tens4}) only 
the usual fields of the string are present.
The Euler-Lagrange equations result from the variation of the
Eq.(\ref{acao1}) together with the conditions for the Nielsen-Olesen
\cite{Nielsen} vortex Eqs.(\ref{vortex1}-\ref{vortex2}), and yield:

\begin{equation}
\begin{array}{cc}
\varphi '' + \frac{1}{r}\varphi ' + \frac{\varphi P^2}{r^2} -
\varphi [\lambda_{\varphi}(\varphi^2 - \eta^2) + 2f_{\varphi \sigma}\sigma^2]=0\\
\sigma'' + \frac{1}{r}\sigma' + \sigma[A^2 +
(f_{\varphi \sigma}\varphi^2 + \lambda_{\sigma} \sigma^2 -
m_{\sigma}^2 + l^2 \Lambda^2)] =0\\
P'' - \frac{1}{r}P' - q^2 \varphi^2P = 0, \hspace{.5 true cm} A'' + \frac{1}{r}A' + e^2 \sigma^2A =0, 
\end{array}\label{equa1}
\end{equation}

\noindent
while the torsion wave equation  is
given by:

\begin{equation} 
\Box_g \Lambda =\frac{l^2}{\alpha} \sigma^2 \Lambda. \label{phi} 
\end{equation}

Above, a prime denotes differentiation with respect to the radial
coordinate $r$. The general
solution for the SCCS will be found in the weak-field approximation
together with junction conditions for the external metric.

\section{The external solution \zero}

Now, we proceed to solve the previous set of equations for an observer
 outside the SCCS stressed by torsion, focusing on
  the external metric which satifies the constraint $r_0\leq r \leq
\infty$.  The external contribution to the energy-momentum of the string reads

\begin{equation} {\cal T}^{\mu}_{\nu} = \frac{1}{4} g^{\mu \alpha} g^{\beta \rho} 
F_{\alpha \beta} F_{\nu \rho } - \delta^{\mu}_{\nu} g^{\sigma \alpha} g^{\beta
\rho} F_{\sigma \beta} F_{\alpha \rho}.\label{stensor1}
\end{equation}

\vspace{.3 true cm}

This tensor is the external energy-momentum tensor of a SCCS with no
torsion. If we observe the asymptotic conditions, Eq.(\ref{config1})
 and Eq.(\ref{config2}), we see that the only field that does not vanish
is the $A_\mu$-field that is responsible for carrying off the string the
effects of the current on it. The torsion contribution to the external 
energy-momentum tensor is given by

\begin{equation}
{\cal T}_{\nu_{tors} }^{\mu} =  2\alpha g^{\mu \alpha}
\partial_{\alpha}\Lambda \partial_{\nu}\Lambda  - \alpha \delta^{\mu}_{\nu }
g^{\alpha \beta} \partial_{\alpha}\Lambda
\partial_{\beta}\Lambda \label{stensor3} . 
\end{equation}

For this configuration, the energy-momentum tensor displays the
following symmetry properties:

\begin{equation} {\cal T}^t_t = -{\cal T}^r_r = {\cal
T}^{\theta}_{\theta}= -{\cal T}^z_z.\label{sim20}  \end{equation}

Then, the only one component of $\Lambda$ in Eq.(\ref{phi}) to be
solved is the $r-$dependent function $\Lambda(r)$. The solution reads:

\begin{equation} \Lambda(r) = \lambda \ln(r/r_0) \label{ssigma}.
\end{equation}

The vacuum solution of Eqs.(\ref{gmg}) are found from the
symmetries (\ref{sim20}). Hence the solutions of $\beta (r) $ and
$\gamma(r) $ are given by

\begin{equation} 
\beta =Br, \hspace{1 true cm}
\gamma = m^2 \ln{r/r_0}.\label{eq2} 
\end{equation}

To find the $\psi$-solution, we can use
the condition:

\begin{equation} R = 2 \Lambda'^2 e^{2(\psi - \gamma)}.\label{esca}
\end{equation}

This condition is different from the usual one\cite{Patrick1}
because the scalar curvature $R $ does not vanish, and
opposedly it is linked to the torsion-field
 $\Lambda$. Then, this condition has the same form as the one for a
SCCS in a scalar-tensor theory \cite{Cris1}. By making use of
solutions (\ref{ssigma}), (\ref{eq2}), we find:

\begin{equation} \psi = n \ln{(r/r_0)} - \ln{\frac{(r/r_0)^{2n} +
k}{(1+k)}}.\label{eq3} \end{equation}

Thus we see that from the solutions of the SCCS Eqs.(\ref{eq2},\ref{eq3}),
there exists a relationship between the parameters $n, \lambda $ and
$m$ given by $n^2 = \lambda^2 + m^2 $.

With the above results, we find that the external metric for the SCCS
takes the form:

\begin{equation}
ds^2 =  \left( \frac{r}{r_0} \right)^{-2n} W^2(r) \left[
\left( \frac{r}{r_0}\right)^{2m^2} (-dt^2 +dr^2) + B^2r^2d\theta^2 \right] + \left( \frac{r}{r_0} \right)^{2n} 
\frac{1}{W^2(r)} dz^2 \label{m8},
\end{equation}

\noindent
with $W(r) = [(r/r_0)^{2n} + k]/[1+k].$

The external
solution alone does not provide a complete description of the physical
situation. We proceed hereafter to find the junction
conditions to the internal metric in order to obtain an appropriate
accounting for the nature of the source and its effects on the surrounding space-time.

\section{SCCS solution: The weak-field approximation \zero}

Nowlet us find the Einstein-Cartan solutions for a SCCS by considering
the weak-field approximation. Thus, the space-time metric may be expanded in terms of a small
parameter $\varepsilon $ about the values $g_{(0)\mu \nu} =
diag(-1,1,1, 1)$, then:

\begin{equation}
g_{\mu \nu} = g_{(0) \mu \nu} + \varepsilon h_{\mu
\nu}, \hspace{1 true cm}
\tilde T_{\mu \nu} = \tilde T_{(0) \mu \nu} +
\varepsilon \tilde T_{(1) \mu \nu}.  \end{equation}

The $\tilde T_{_{(0)}\mu \nu}$ tensor corresponds to the
energy-momentum tensor in a space-time with no curvarture.
However, torsion is embeeded. $ \tilde T_{_{(1)}\mu \nu}$
represents the part of the energy-momentum tensor containing
curvature and torsion.  Next we proceed to
define some important quantities useful for the analysis to come.

The energy-momentum density  and tension of the thin SCCS are given by:

\begin{equation} U = -2 \pi \int_{0}^{r_0}\tilde T^t_{_{(0)}t} r dr; \hspace{1 true cm} T = - 2 \pi \int_{0}^{r_0} \tilde T^z_{_{(0)}z} rdr \label{U} \end{equation}

The remaining components follows as

\begin{equation} X = -2 \pi \int_{0}^{r_0} \tilde T^r_{_{(0)}r} r dr;  \hspace{1 true cm} Y = -2 \pi \int_{0}^{r_0} \tilde
T^{\theta}_{_{(0)}\theta} r dr.\label{X} 
\end{equation}

The energy conservation in the weak-field approximation, reduces
to

\begin{equation} r\frac{d \tilde T^r_{_{(0)}r}}{dr} = (
\tilde T^{\theta}_{_{(0)}\theta} - \tilde T^r_{_{(0)}r}),\label{bian2}
\end{equation}

\noindent
where $ \tilde T_{_{(0)} \mu \nu}$ represents the trace of the energy-momentum tensor with tprsion.

For computing the overall metric, we use the Einstein-Cartan in the
quasi-Einsteinian Eq.(\ref{gmg}), where it gets the form $G^{\mu
\nu}(\{\}) = 8 \pi G \tilde T^{\mu \nu}_{_{(0)}}$ in the
weak-field approximation, with the tensor $\tilde T_{_{(0)}\mu \nu}$
(being first order in $G$) containing torsion. After integration we have:

\begin{equation} \int_0^{r_0} r dr (\tilde T^{\theta}_{_{(0)} \theta}+
\tilde T^r_{_{(0)}r})= r_0^2 \tilde T^r_{_{(0)}r}(r_0) = r_0^2\left[
\frac{A'^2(r_0)}{2 e^2} + \frac{\alpha}{2} \Lambda'^2(r_0)\right] . 
\label{ap-1} \end{equation}

To find the internal energy-momentum tensor, it is more convenient to
use Cartesian coordinates\cite{Patrick1}.  For this purpose we
proceed to calculate the cross-section integrals of $\tilde
T^x_{_{(0)}x}$ and $\tilde T^y_{_{(0)}y}$ that in
cartesian coordinates reads

\begin{equation} 
\begin{array}{ll}
\tilde T^{x}_{_{(0)}x} = c[\varphi'^2 +
\sigma'^2 + \left(\frac{A'}{e}\right)^2 + \alpha \Lambda'^2] + s\frac{\varphi^2P^2}{r^2} +
\frac{1}{2}\left(\frac{P'}{qr}\right)^2 -\frac{1}{2} \sigma^2 A^2 -
2V \\
\tilde T^{y}_{_{(0)}y} = s [\varphi'^2 +
\sigma'^2 + \left(\frac{A'}{e}\right)^2 + \alpha \Lambda'^2] + c\frac{\varphi^2 P^2}{r^2} +
\frac{1}{2}\left(\frac{P'}{q r}\right)^2 - \frac{1}{2}\sigma^2 A^2
- 2V, \label{tens31} 
\end{array}
\end{equation}

\noindent
where $c=\cos^2 \theta -\frac{1}{2}$ and $s=\sin^2 \theta
-\frac{1}{2}$.
This way we found:

\begin{equation} \int rdrd\theta \tilde T^x_{_{(0)}x}=\int rdrd\theta
\tilde T^y_{_{(0)}y}= \pi \int r dr[\left(\frac{P'}{q r}\right)^2
- \sigma^2 A^2 - V] = - W.  \end{equation}

Using the fact that
$\tilde T^r_{_{(0)}r} + \tilde
T^{\theta}_{_{(0)}\theta} = \tilde T^x_{_{(0)}x} + \tilde
T^y_{_{(0)}y}$, then we have:

\begin{equation} X + Y = 2 W =-2 \pi r_0^2\left[
\frac{A'^2(r_0)}{e^2} + \alpha \Lambda'^2(r_0)\right] ,
\end{equation}

\noindent
which can be computed by integration of Eq.(\ref{equa1})

\begin{equation}
A'(r) = \frac{e J}{\sqrt{2}\pi r}, \hspace{1 true cm}
J= \sqrt{2} \pi e\int_0^{r_0} r dr \sigma^2 A,
\end{equation}

\noindent
where $ J$ is the current density. Thus, the torsion density can be computed by integration of Eq.(\ref{phi})

\begin{equation}
\Lambda'= \frac{S}{\sqrt{2}\pi \alpha r},\hspace{1 true cm} S =\sqrt{2}\pi  l^2\int^{r_0}_{0} r dr
\sigma^2 \Lambda, \hspace{1 true cm} \label{s2}
\end{equation}

\noindent
where $S$ is the torsion density. With these considerations we found the string structure. Then, we obtain

\begin{equation} 
W  = -\frac{1}{2\pi }\left(J^2 + \nu S^2\right).  
\end{equation}

\noindent
with $\nu = 1/\alpha $.

In addition, we can assume that the string is infinitely thin so that its
stress-energy tensor is given by

\begin{equation} \tilde T^{\mu \nu}_{string} = diag [U , -W, -W,
-T]\delta(x)\delta(y).\label{stensor2} \end{equation}

It worths to note that definitions for both string energy $U$ and
tension $T$, as in equations (\ref{U}), already incorporate
information on the torsion.

 By virtue of the presence of the external current we use the form
Eq.(\ref{stensor2}) for the string energy-momentum tensor as well
as Eq.(\ref{stensor1}) and Eq.(\ref{stensor3}) for the external
energy-momentum tensor in linearized solution to zeroth order in
G. In the sense of distributions we have, $\nabla^2 ln (r/r_0) = 2 \pi
\delta(x) \delta(y)$, $ \nabla^2 (\ln (r/r_0))^2 = 2/r^2$ and $
 \nabla^2(r^2\partial_i \partial_jln(r/r_0))=4\partial_i \partial_j ln(r/r_0)$.

The energy-momentum tensor of the string source $\tilde T_{(0)\mu
\nu}$, in Cartesian coordinates, possesses no curvature, which is the
well-known result \cite{Patrick1,Cris1}, but
 does have torsion which produces the following energy-momentun tensor

\begin{equation}
\begin{array}{ll} 
\tilde T_{(0) tt} = U \delta(x)\delta(y) + \frac{(J^2
+ \nu S^2)}{4\pi}\nabla^2 \left(ln\frac{r}{r_0}\right)^2 ,\\
\tilde T_{(0) zz} = -T \delta(x)\delta(y) + \frac{(J^2
- \nu S^2)}{4\pi}\nabla^2\left(ln\frac{r}{r_0}\right)^2,\\
\tilde T_{(0) ij} = (J^2 +\rho S^2) \delta_{ij}\delta(x)\delta(y)
- \frac{ (J^2 + \nu S^2) }{2 \pi}\partial_i \partial_j ln(r/r_0),
\end{array}
\end{equation}

\noindent
where the trace is given by $\tilde T_{(0)}= - (U + T - J^2 -\nu S^2 )\delta(x)
\delta(y) - \frac{\nu S^2}{2 \pi}
\nabla^2\left(\ln\frac{r}{r_0}\right)^2 $ .

Now let us find the matching conditions to the external solution. For
this purpose, we shall use the linearized Einstein-Cartan equation in the form

\begin{equation} \nabla h_{\mu\nu} =-16 \pi G (\tilde T_{(0)\mu \nu} -
\frac{1}{2} g_{_{(0)} \mu \nu} \tilde T_{(0)}).\label{ricci1}
\end{equation}

The internal solution to equation (\ref{ricci1}) with source yields:

\begin{equation} 
\begin{array}{cc}
h_{tt} =-4 G [J^2(\ln(r/r_0))^2 + (U-T
+J^2 + \nu S^2)\ln(r/r_0)] \\
h_{zz} =-4 G [J^2 \ln(r/r_0))^2 + (U-T
-J^2 - \nu S^2)\ln(r/r_0)] \\
h_{ij} = -2 G (J^2 + \nu S^2) r^2 \partial_i
\partial_j \ln(r/r_0) - 4 G \delta_{ij}\left[\left( U + T +J^2 + \nu S^2) \ln(r/r_0) \right) +
S^2\left(\ln \frac{r}{r_0}\right)^2\right].
\end{array}
\end{equation}

This corresponds to the solution in Cartesian coordinates. We note
that the torsion appears explicitly in the transverse components of the
metric. To analyse the solution for the junction condition to the external
 metric let us transform it back into cylindrical coordinates.

\section{Matching Conditions \zero }

It is possible to find the matching conditions \cite{Kopczynski} to
the external solution. In the case of a space-time with torsion we
can find the junction conditions using the fact
 that $[\{^\alpha_{\mu\nu}\}]_{_{r=r_0}}^{(+)}=
[\{^\alpha_{\mu\nu}\}]_{_{r=r_0}}^{(-)}$, and the metricity condition
$[\nabla_{\rho}g_{\mu \nu}]_{_{r=r_0}}^{+}= [\nabla_{\rho}g_{\mu
\nu}]_{_{r=r_0}}^{-}= 0$, to find the continuity conditions

\begin{eqnarray} 
&[g_{\mu \nu}]_{_{r=r_0}}^{(-)} =
[g_{\mu \nu}]_{_{r=r_0}}^{(+)}, \nonumber \\ 
&[\frac{\partial g_{\mu
\nu}}{\partial x^{\alpha }}]_{_{r=r_0}}^{(+)} + 2 [g_{\alpha \rho}
K_{(\mu \nu)}^{\hspace{.3 true cm} \rho}]_{_{r=r_0}}^{(+)} =
[\frac{\partial g_{\mu \nu}}{\partial x^{\alpha}}]_{_{r=r_0}}^{(-)} +
2 [g_{\alpha \rho} K_{(\mu \nu)}^{\hspace{.3 true cm}
\rho}]_{_{r=r_0}}^{(-)} \label{junc1} 
\end{eqnarray}

\noindent
where $(-)$ represents the internal region and $(+)$ corresponds the
external region around $r = r_0$. In analysing the junction conditions
we notice that the contortion contributions do not appear neither in
the internal nor in the external regions\cite{Kopczynski,Volterra}.

To match our solution with the external metric we used the metric in
cylindrical coordinates,
 which is obtained from the coordinate transformations:

\begin{equation} r^2 \partial_i \partial_j ln(r/r_0)dx^idx^j =
r^2d\theta^2 - dr^2, \end{equation}

\noindent
Unfortunately, for this goal we cannot use the metric
the way it stands. Therefore, we have to change the
radial coordinate to $\rho$, using the constraint (symmetry) $g_{\rho
\rho }= -g_{tt}$, to have, to first order in $G$, $\rho=r[1+ a_1-a_2\ln(r/r_0) - a_3(\ln(r/r_0)^2]$.

In this case we have $a_1=G(4U + J^2 + \nu S^2 )$, $a_2= 4GU$ and $a_3=-2G(J^2+ \nu S^2)$, which corresponds to
the magnetic configuration of the string fields\cite{Patrick1}. The transformed metric yields:

\begin{equation} 
\begin{array}{cc}
g_{tt} = -\{1 + 4G[J^2 (\ln(\rho/r_0))^2 + (U - T +
J^2 + \nu S^2)\ln(\rho/r_0)]\} = -g_{\rho \rho}\\
g_{zz} = \{1 - 4G[ (J^2 + \nu S^2)(\ln(\rho/r_0))^2 + (U - T +
J^2- \nu S^2)\ln(\rho/r_0)]\}\\
g_{\theta \theta} = \rho^2\{1 - 8G(U + \frac{(J^2 + \rho S^2)}{2}) +
 4G(U - T - J^2 - \rho S^2)\ln(\rho/r_0)] + 4GJ^2(\ln(\rho/r_0))^2\}\label{m5}.  
\end{array}
\end{equation}

Now we can find the external parameters $B $, $n$ and $m$ as
functions of the source structure. If we consider the junction of
the equation (\ref{junc1}),  after the linearization, and using
the limit $|n\ln(\rho/r_0)|<<1$, we have:

\begin{equation} 
\begin{array}{ll}
n\left(\frac{1-k}{1+k}\right) = 2G\left( U - T - J^2 - \nu S^2
\right)\\
B^2= 1-8G\left(U + \frac{(J^2 + \rho S^2)}{2}\right)\\
m^2= 4G(J^2 + \nu S^2).  
\end{array}
\end{equation}

\noindent
and using the derivative of the expression Eq.(\ref{ssigma}) and the Eq.(\ref{s2}), we find 

\begin{equation}
\lambda = \frac{\nu}{\sqrt{2}\pi }S.
\end{equation}

This expression completes the derivation of the full metric
components.
In analysing the metric of the SCCS with torsion we note that the
contribution of torsion appears in the $\theta \theta$-metric component,
which is important in astrophysical applications such as gravitational
lensing studies because this component is
linked to the deficit angle. Next we present two preliminary applications of the formalism here
introduced assuming that such a kind of torsioned SCCSs really exist.
Firstly, we focus on the issue of the deviation of a  particle moving
near the string, and later on we attempt to perform a comparative analysis of  the observations performed by the COBE satellite, with the effects this sort of string may produce on the CMBR, supposed to interact
with it as discussed in this paper.

\section{Particle deflection near a torsion SCCS \zero}

We know that when the string possesses current there appear
gravitational forces.  We shall consider the effect that torsion
plays on the gravitational force generated by SCCS on a particle
moving around the defect, initially with no charge. We consider the particle speed $|{\bf v}| \leq 1 $, condition under which
the geodesic equation becomes:

\begin{equation} \frac{d^2x^i}{d\tau^2} + \Gamma_{tt}^{i} =0,
\end{equation}

\noindent
where $i$ is the spatial coordinate and the connection can be written as
Eq.(\ref{kont1})), in this manner the gravitational force of the string (per unit
length) gets the form

\begin{equation} F_G = \frac{1}{2}( \nabla h_{tt} - \frac{S}{\rho}),
\end{equation}

\noindent
with $g_{tt}= -1- h_{tt} $ in Eq.(\ref{m5}). 
We also note that the gravitational force is related to the $h_{tt}$ component
that has no explicit dependence on the torsion. From the last equation, the  force the SCCS exerts on a test particle can be explicitly written as

\begin{equation} F_G = -\frac{1}{\rho} \left[ 2GJ^2\left(1 + \frac{(U-
T + \nu S^2 )}{J^2} + 2\ln(\rho/r_0) \right) + S\right]. \label{force}
\end{equation}

A quick glance at the last equation allows us to understand the essential role 
torsion may in the context of the present formalism. As we show below, this extra-term  yields an amplification  of the total force a particle close to the SCCS will undergo.

If torsion is
present, even in the case the string has no current, an attractive
gravitational force appears. In the context of the SCCS torsion acts
as a enhancer of the force a test particle feels outside the string.
 In our summary,
we discuss a bit further potential applications of this new result
to astrophysics and cosmology.

\section{Angular deficit and COBE map \zero}

Recently many works  \cite{Palle,Garcia3} have
shown that the COBE data are compatible with  Einstein-Cartan gravity.
 In this section, we analyse the effects of a screwed superconducting cosmic string on
the primordial microwave background radiation using the COBE data. To
this end, we need to compute the angular deficit introduced by
torsion. The hidden idea here is that the cosmic large-scale density
fluctuations could have had origin during the appearance of Cosmic
String defects or due to interaction with them during the late stages
of the Universe's evolution. The torsion would modify properties of light and
radiation interacting with a cosmic string pervaded by screw
dislocations in such a way that the density fluctuations induced might match
those ones measured by COBE\cite{smoot99}. The DMR (Differential Microwave Radiometer) instrument of COBE has
provided temperature sky maps leading to the rms sky variation where
the beam separation in the COBE experiment is $\theta_1 - \theta_2 =
60^o$ .

Each string that affects the photon beam induces a temperature
variation \cite{Kaiser,Stebbins}, of the order of magnitude as 
$\frac{\delta T}{T} \sim \delta \leq 10^{-6} ({\rm COBE})$, 
where $\delta$ is the angular deficit.  If we consider the metric
Eqs.(\ref{m5}), projected into the space-time perpendicular
to the string, i. e., $dz=0$, then we have:

\begin{equation} ds^2_{\perp}= (1-h_{tt})[-dt^2 + dr^2 + (1- b)
r^2d\theta^2], \end{equation}

with $h_{tt}$ given by Eq.(\ref{m5}), and $b$ calculated from junction
conditions using $\beta^2= (1-b)\rho^2$.

Then, in first order in $G$, the deficit angle gets:

\begin{equation}
\delta= b\pi =8\pi G\{U+\frac{J^2(1+ \ln(\rho/r_0))}{2} + \frac{1}{2}\nu S^2(1 - 2\ln(\rho/r_0))\}.
\end{equation}

We can interpret this angular deficit $\delta$ as being due to three
different contributions: $\delta_s$ to an ordinary cosmic string,
$\delta_J$ to the current
 and $\delta_{_{tors}} $ to the torsion field, respectively. In the case
 of the ordinary cosmic string the angular deficit is given
by $\delta_s = 8 \pi G U$, which in this work corresponds to the
case where both current $J$ and torsion $S$ vanish. In this
situation\cite{Leandros}, it is demonstrated that cosmic string models
are more consistent with the COBE data \cite{Bennett} for a wider
range of cosmological parameters than the standard CDM models, and the
numerical simulations have confirmed these predictions \cite{Bouchet}.

When the cosmic string carries current, we have used results of
Ref.\cite{Babul2} for the current, that is, a configuration with the
maximum current $J \sim \eta $, for $\eta = 10^{16} GeV$ as well-known
for grand unification theories. In such a case, we found $\delta_{J} = 8\pi
GJ^2 \sim 10^{-6}$ or less what is compatible with COBE data. As it is
easy to see, we neglected the logarithmic term because we consider the
experiment is being performed close to the string surroundings.

However, in the situation where the cosmic string is stressed by
torsion, the issue is more difficult because we have no idea
about the energy density torsion puts in the Universe via cosmic strings.
Therefore, if a cosmic string actually formed (and it is a good mechanism
to generate density fluctuations that can be measured by COBE),
then we can estimate the density of torsion the string induces in
the cosmic background. To this purpose we choose the value $\nu \sim 10^{38}
GeV^{2} \sim 1/G$; in this case, we have the torsion
energy density $S \sim 10^{-3}$ with $\delta_{tors}= 8\pi G \nu S^2$.
As one can check by substituting in the previous section, the inferred value 
for $S$ enlarges the intensity of the net force undergone by a test particle 
encircling the SCCS.

\section{Summary}

It is possible for torsion to have had a physically relevant role
during the early stages of the Universe's evolution. Along these lines,
torsion fields may be potential sources of dynamical stresses which,
when coupled to other fundamental fields (i. e., the gravitational
field), might have performed an important action during the phase
transitions leading to formation of topological defects such as the
SCCSs here we focused on. It therefore seems a crucial issue to
investigate basic models and scenarios involving cosmic defects within
the torsion context. We showed that in this picture there exists the
possibility for SCCSs to effect the spectrum of primordial density
perturbations, whose imprints could be seen in the relic cosmic
microwave background radiation as observed by COBE.

We also showed that torsion has a non-negligible contribution to
the geodesic equation obtained from the contortion term. From a
physical point of view, this contribution is responsible for the
appearance of a stronger attractive force acting on a test-particle. Using
the COBE  data, we found that the torsion density contribution $S$
is the order of $\sim 10^{-3}$. If we compute the force strength,
Eq.(\ref{force}), in association with the above estimative and
data coming from COBE observations, we can show that the torsion
contribution to this force is $10^{3}$ times bigger than the
corresponding to a current-carrying string compared to the one induced
by the gravitational interaction itself.

This peculiar fact may have meaningful astrophysical and
cosmological effects. Let us imagine for a while a compact object
(CO): a black hole or an exotic cosmic relic such as a boson,
strange  mirror star, for instance, orbiting around the SCCS.
Because the acceleration induced on the radial component of its
orbital motion is about one thousand stronger than in ordinary
cases, then, we can expect the changes it provokes in the
quadrupole moment of the system (SCCS + CO) to be enhanced by a
large factor so that the gravitational wave (GW) signal expected
from the CO inspiraling onto the SCCS could be above the lower
strain sensitivity threshold of planned LIGO, VIRGO, GEO-600, etc.
interferometric GW observatories, for distances even as the Hubble
radius. Moreover, this very strong force may also turn the SCCS a
potential source of hard X-ray and $\gamma$-ray transient emissions. These
radiations can be emitted by matter (primordial gas and/or dust
clouds, or something else) accreting onto the SCCS as the material
gets closer and becomes heated due to the powerful tidal
stripping. All these issues, we plan to address to in a forthcoming
work including the Sachs-Wolfe effect in space-time with torsion \cite{Herman1}.

\vspace{1 true cm}

{\bf Acknowledgments:} We would like to express our deep gratitude to
Prof.J.A. Helay\"el-Neto and Prof. 
V.B. Bezerra for helpful discussions on the subject of this paper. 
C.N Ferreira and H.J.
Mosquera Cuesta (CLAF) would like to thank (CNPq-Brasil) for 
financial support. We also thank Centro 
Brasileiro de Pesquisas F\ii sicas (CBPF) and the Abdus Salam 
ICTP-Trieste for hospitality
during the preparation this work. Garcia de Andrade thanks (FAPESP)
and (CNPq-Brasil) for financial support and Dr. Rudnei Ramos, Prof.
I. Shapiro for important suggestions concerning this work.

\end{document}